\documentclass[11pt,a4paper]{article}
\usepackage[a4paper,top=4cm,bottom=4cm,left=3cm,right=3cm]{geometry}
\usepackage{amsmath, amsthm, amssymb, amsfonts, mathtools, bbold, pgfplots}
\usepackage{hyperref}
\usepackage{tikz}
\usepackage{listings}
\usetikzlibrary{shapes.geometric, arrows}
\usetikzlibrary{positioning}
\tikzstyle{doubleRECT} = [rectangle, rounded corners, minimum width=3cm, minimum height=2cm, text centered, draw=black, fill=white!30]
\tikzstyle{line} = [draw, -latex']

\title{A new ECDLP-based PoW model}
\author{Alessio Meneghetti, Massimiliano Sala, Daniele Taufer}
\date{\today}

\theoremstyle{plain}
\newtheorem{thm}{Theorem}[section]

\newtheorem{rmk}[thm]{Remark}

\renewcommand{\H}{\mathcal{H}}

\makeatletter
\let\@@citation@@=\citation

\renewcommand{\citation}[1]{\@@citation@@{#1}%
\@for\@tempa:=#1\do{\@ifundefined{cit@\@tempa}%
  {\global\@namedef{cit@\@tempa}{}}{}}%
}

\def\@lbibitem[#1]#2#3\par{%
  \@ifundefined{cit@#2}{}{\item[\@biblabel{#1}\hfill]}%
  \if@filesw
      {\let\protect\noexpand
       \immediate
       \write\@auxout{\string\bibcite{#2}{#1}}}\fi\ignorespaces
  \@ifundefined{cit@#2}{}{#3}}
\def\@bibitem#1#2\par{%
  \@ifundefined{cit@#1}{}{\item}%
  \if@filesw \immediate\write\@auxout
    {\string\bibcite{#1}{\the\value{\@listctr}}}\fi\ignorespaces
  \@ifundefined{cit@#1}{}{#2}}
\makeatother

\begin{document}
\maketitle

\begin{abstract}
We lay the foundations for a blockchain scheme, whose consensus is reached via a proof of work algorithm based on the solution of consecutive discrete logarithm problems over the point group of elliptic curves.
In the considered architecture, the curves are pseudorandomly determined by block creators, chosen to be cryptographically secure and changed every epoch.
Given the current state of the chain and a prescribed set of transactions, the curve selection is fully rigid, therefore trust is needed neither in miners nor in the scheme proposers.
\end{abstract}

\vspace{0.5cm}



\section{Introduction}

A \emph{proof of work} (PoW) is a procedure that allows a prover to demonstrate that he is very likely to having performed a specific amount of computational work within a prescribed interval of time \cite{SurveyPoW}.

This concept has been formalized in 1999 \cite{PoWandBP}, although previous instances of delaying functions conceived for similar purposes had appeared earlier \cite{Benchmark2, Benchmark1, JunkEmails, Metering, Lotteries, ClientPuzzles, TimeLock}. 

Since 2008, PoW-methods have been attracting a considerable interest as Bitcoin \cite{Btc} introduced a PoW-based consensus algorithm, which puts miners in competition for solving a cryptographic challenge.
Bitcoin's consensus relies on a hashcash system \cite{HashcashFirst, Hashcash}, whose workload may be easily adjusted with a fastly verifiable output.
Despite their high efficiency and easy implementation, all the hashcash-based protocols share a common limitation: the huge amount of computations employed by nodes becomes useless after the consensus is reached.
This aspect has been raising environmental concerns and many solutions have been proposed to reduce these energy-intensive computer calculations.

A promising countermeasure to this issue is the adoption of \emph{bread pudding protocols} \cite{PoWandBP}. They face the aforementioned problem by performing a computational work that is reusable either for practical \cite{CureCoin, rPoW, Permacoin, PoX}, cryptographical \cite{PoWandBP, MicroMint} or mathematical \cite{Primecoin} reasons.
Moreover, the latter class of systems encloses several protocols that are meant to be research propellants \cite{uPoW}, namely designed to boost the commitment upon the solution of difficult mathematical problems.

Along the same line, we have proposed \cite{NewECDLP} a blockchain architecture with a PoW-consensus algorithm based on the solution of the \emph{Discrete Logarithm Problem} over the point groups of elliptic curves (ECDLP).
In this work, we provide that germinal proposal with precise mathematical foundations and further implementation details.

The idea of basing the PoW on ECDLP has already appeared in other works \cite{DLOG, CryptoTool}, as this problem is widely studied and applied in cryptographic protocols.
However, the considered curves does not usually fulfil the standard security criteria \cite{SecCrit}, especially for what concerns the \emph{fully rigidity}: the network has to initially trust an authority that is providing the curve parameters.

In this work we radically solve this issue by designing a PoW-system based on elliptic curves that are changing over the time.
Since the curves are pseudo-randomly constructed and satisfy general security conditions, a malicious user could attack the chain only by breaking the ECDLP for an immense class of elliptic curves, which is currently considered infeasible.
									
This paper is organized as follows: after a quick summary of the ECDLP in Section \ref{sec:ECDLP}, we deline the proposed blockchain architecture in Section \ref{sec:Architecture} and its blocks construction in Section \ref{SB} and \ref{EB}.
The strong points of this system are discussed in Section \ref{sec:Discussion}, including a theorem on the security of our system, while in Section \ref{sec:Conclusion} future work directions are suggested.

\section{ECDLP} \label{sec:ECDLP}

The ECDLP is a renown problem that consists of finding an integer $N \in \mathbb{N}$ such that the $N$-th multiple of a \emph{base point} $P$ of an elliptic curve $E$ over a finite field equals another given point $Q$, i.e. $Q = N \cdot P$.

Here we are only interested in elliptic curves over prime fields $\mathbb{F}_p$ and determined by their short Weierstrass equation $y^2 = x^3 + Ax + B$.
Solving ECDLP for a curve $E$ over large fields is considered to be a difficult challenge except for degenerate cases.

\subsection{The general case}

Currently the best known \emph{general} attacks are Baby-Step Giant-Step \cite{Shanks} and Pollard's Rho - Kangaroo algorithms \cite{Pollard}, which have an asymptotic complexity of $O(\sqrt{|E|})$, where $|E|$ is the size of $E$.
These are general parallel collision-finding algorithms, which work over any groups, i.e. no properties of the underlying structure but the operation definition are used.

The introduction of Semaev's polynomials \cite{SemaevSum} have suggested the existence of subexponential algorithms to solve ECDLP, however no clear evidence has emerged.
Pairings-based attacks \cite{FR, MOV}, Index calculus \cite{AmadPintSala, NewIndexC, IndexC} and Xedni calculus \cite{Xedni} have been recently being studied, but none of them seem to significantly reduce the problem complexity of the general case, so far.

\subsection{Special cases}

There are some families of curves whose ECDLP is known to be easier than the general case, namely there are algorithm for efficiently solving it. Consequently, these curves have to be carefully avoided for designing a ECDLP-based protocol.
The following is a concise summary of those particular attacks, the curve on which they may be efficiently applied and how we avoid them.

\begin{small}
\begin{center}
\begin{tabular}{ |c|c|c|c| } 
 \hline
Attack & It applies on curves & To avoid it: use & Ref. \\
 \hline
 Weil-descent & over composite fields & prime fields & \cite{WeilD0, WeilD} \\ 
 Polig-Hellman & of composite orders & prime orders & \cite{PH} \\ 
 Semaev, Satoh-Araki, Smart & anomalous & non-anomalous curves & \cite{Semaev, SatAr, Smart} \\
 Menezes-Okamoto-Vanstone & low embedding degree & high embedding degree & \cite{MOV} \\
 Frey-R\"uck & low embedding degree & high embedding degree & \cite{FR} \\
 Wiener-Zuccherato & low CM discriminant & high CM discriminant & \cite{CMDisc} \\
 \hline
\end{tabular}
\end{center}
\end{small}

\section{A sample blockchain architecture} \label{sec:Architecture}

To show how our PoW works, we introduce a schematic sample ledger architecture, but our algorithm may easily be adapted for any blockchain scheme.
Our architecture is based on two types of blocks:
\begin{enumerate}
\item[{[EB]}] An \emph{Epoch Block} contains, aside from the header and a list of transactions, a prime number $p$, an elliptic curve $E$ defined over $\mathbb{F}_p$ and a base point $P$ of $E$, all to be determined by the proposing miner. 

Moreover, it encloses as PoW two integer $N_1$ and $N_2 \in \{0, \dots, |E| - 1\}$ to be discovered by the proposing miner such that $N_i \cdot P$ are points of $E$ deterministically determined from the header of the block.

These EBs occur once every $2016$ blocks in the blockchain.

\item[{[SB]}] The \emph{Standard Blocks} are just a light version of the EB blocks, they are constructed in the same way except for $p$, $E$, $P$, which are inherited from the last EB block of the chain. 
 
SBs constitute the vast majority of the blocks of the chain.
\end{enumerate}

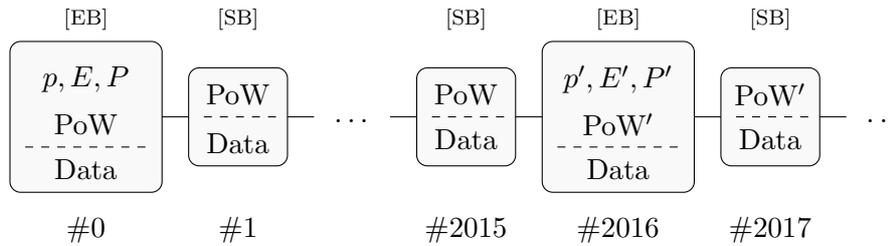
\begin{center}
\vspace{0.2cm}
\begin{tikzpicture}
// BLOCKS
\node (b0) [draw = black, rectangle, rounded corners, fill=gray!5, minimum width=2cm, minimum height=2cm,] at (0,0) {};
\node (Name) [above of = b0, node distance=1.3cm] {\scriptsize [EB]};
\node (pEP) [below of = b0, node distance=-0.5cm] {$p, E, P$};
\node (lev) [below of = b0, node distance=0.1cm] {PoW};
\draw [dashed] (-0.8,-0.4) -- (0.8,-0.4);
\node [below of = lev, node distance=0.6cm] {Data};
\node [below of = b0, node distance=1.5cm] {\#0};

\node (b1) [draw = black, rectangle, rounded corners, fill=gray!5, minimum width=1.3cm, minimum height=1.3cm,] at (2,0) {};
\node (Name) [above of = b1, node distance=1.3cm] {\scriptsize [SB]};
\node (lev1) [below of = b1, node distance=-0.3cm] {PoW};
\draw [dashed] (1.55,0) -- (2.5,0);
\node [below of = lev1, node distance=0.65cm] {Data};
\node [below of = b1, node distance=1.5cm] {\#1};

\node (b2015) [draw = black, rectangle, rounded corners, fill=gray!5, minimum width=1.3cm, minimum height=1.3cm,] at (5,0) {};
\node (Name) [above of = b2015, node distance=1.3cm] {\scriptsize [SB]};
\node (lev2015) [below of = b2015, node distance=-0.3cm] {PoW};
\draw [dashed] (4.55,0) -- (5.5,0);
\node [below of = lev2015, node distance=0.6cm] {Data};
\node [below of = b2015, node distance=1.5cm] {\#2015};

\node (b2016) [draw = black, rectangle, rounded corners, fill=gray!5, minimum width=2cm, minimum height=2cm,] at (7,0) {};
\node (Name) [above of = b2016, node distance=1.3cm] {\scriptsize [EB]};
\node (pEP) [below of = b2016, node distance=-0.5cm] {$p', E', P'$};
\node (lev) [below of = b2016, node distance=0.1cm] {PoW$'$};
\draw [dashed] (6.2,-0.4) -- (7.8,-0.4);
\node [below of = lev, node distance=0.6cm] {Data};
\node [below of = b2016, node distance=1.5cm] {\#2016};

\node (b2017) [draw = black, rectangle, rounded corners, fill=gray!5, minimum width=1.3cm, minimum height=1.3cm,] at (9,0) {};
\node (Name) [above of = b2017, node distance=1.3cm] {\scriptsize [SB]};
\node (lev2017) [below of = b2017, node distance=-0.3cm] {PoW$'$};
\draw [dashed] (8.55,0) -- (9.5,0);
\node [below of = lev2017, node distance=0.6cm] {Data};
\node [below of = b2017, node distance=1.5cm] {\#2017};

//LINES
\draw (1,0) -- (1.35,0);
\draw (2.65,0) -- (3,0);
\node (dots) at (3.52,-0.05) {\dots};
\draw (4,0) -- (4.35,0);
\draw (5.65,0) -- (6,0);
\draw (8,0) -- (8.35,0);
\draw (9.65,0) -- (10,0);
\node (dots) at (10.52,-0.05) {\dots};
\end{tikzpicture}
\end{center}

EBs basically define the setting (curves and base points) on which the discrete logarithm PoWs will have to be solved in the following epoch. They are slightly heavier to be produced and verified but occur rarely (roughly once every two weeks with a BTC-like difficulty adjustment).



In order to give the specifications of our blocks we need a deterministic function \verb|P_Gen| to construct a point on a given elliptic curve $E$ from a prescribed hash digest $h$, which we treat as an integer for simplicity. The following is a concrete example of such a function.

\begin{lstlisting}
function P_Gen(h,E)
  i = 0
  while #{points of E with x-coord = h + i} = 0:
    i = i + 1
  P = (h + i, *) point of E with 0 $\leq$ * < p/2
  return P
\end{lstlisting}

We notice that the points determined by the above function are affine by construction.
The hash $\H$ that we propose to use in the following is SHA3-512 \cite{Keccak}, which provides a satisfying collision resistance even against post-quantum attacks, but one might conceivably replace it with another properly constructed one.

We also assume that all proposing miners use prescribed signature algorithms and we denote with $\sigma_k(m)$ the signature of the string $m$ obtained by the miner with signing key $k$.

\subsection{Standard Blocks} \label{SB}

A minimal model of a SB consists of a list of valid transactions and a header, which comprises their Merkel root $\mathcal{M}$, the hash of the previous header $h_{\text{prev}}$ and a pair of integers $(N_1,N_2)$ solving 
\begin{equation*}
\textnormal{PoW} : \quad \begin{cases} \verb|P_Gen|(\H(h_{\text{prev}}), E) = N_1 \cdot P,\\
\verb|P_Gen|(\H(\mathcal{M}), E) = N_2 \cdot P.
\end{cases}
\end{equation*}
where $E$ and $P$ are defined in the last EB.

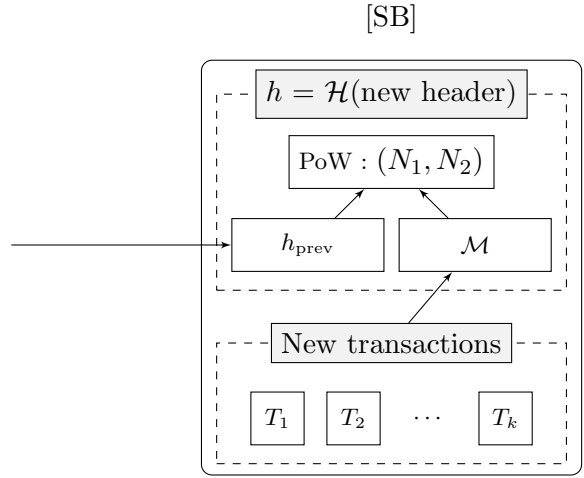
\begin{center}
\begin{tikzpicture}
\node (start) [rectangle, rounded corners, minimum width=5cm, minimum height=5.5cm, text centered, draw=black, fill=white!30] { };
\node [above of = start, node distance=+3.3cm] {[SB]};

\draw [dashed] (-2.3,2.3) rectangle ++(4.6,-2.6);
\node (NewH) [draw = black, rectangle, above of = start, node distance=1.8cm, fill=gray!10] at (0,0.5) {$h$ = $\mathcal{H}$(new header)};

\node (PoW) [draw = black, rectangle, minimum width=2cm, minimum height=0.7cm, above of = start, node distance=1cm] at (0,0.4) {\footnotesize PoW : \normalsize $(N_1, N_2)$};
\node (Prev) [draw = black, rectangle, minimum width=2cm, minimum height=0.7cm, above of = start, node distance=1cm] at (-1.1,-0.7) {\footnotesize $h_{\text{prev}}$};
\node (Merkle) [draw = black, rectangle, minimum width=2cm, minimum height=0.7cm, right of = Prev, node distance=2.2cm] {\footnotesize $\mathcal{M}$};

\draw [dashed] (-2.3,-1.0) rectangle ++(4.6,-1.6);
\node (NewT) [draw = black, rectangle, below of = start, node distance=1.2cm, fill=gray!10] at (0,0.2) {New transactions};
\node (T1) [draw = black, rectangle, minimum width=0.7cm, minimum height=0.7cm, below of = (NewT), node distance=2.5cm] at (-1.5,0.5) {\footnotesize $T_1$};
\node (T2) [draw = black, rectangle, minimum width=0.7cm, minimum height=0.7cm, below of = (NewT), node distance=2.5cm, node distance=2.5cm] at (-0.5,0.5) {\footnotesize $T_2$};
\node (T3)[rectangle, minimum width=0.7cm, minimum height=0.7cm, below of = (NewT), node distance=2.5cm] at (0.5,0.5) {\footnotesize $\cdots$};
\node [draw = black, rectangle, minimum width=0.7cm, minimum height=0.7cm, below of = (NewT), node distance=2.5cm] at (1.5,0.5) {\footnotesize $T_k$};

\path [line] (Merkle) -- node [midway,left] {} (PoW);
\path [line] (Prev) -- node [midway,left] {} (PoW);
\path [line] (NewT) -- node [midway,left] {} (Merkle);
\path [line] (-5,0.3) -- node [midway,above,xshift=-6.5mm] {} (Prev);
\end{tikzpicture}
\end{center} 

\subsection{Epoch Blocks} \label{EB}

An EB is a thick version of a SB, namely it is constructed in a similar fashion but it enclodes three additional data: the prime $p$, the elliptic curve $E$ over $\mathbb{F}_p$ and the base point $P$ of $E$.
\begin{itemize}
\item Generating $p$

The prime number $p$ is the responsible of the expected run time of the PoW. Its size is determined by the difficulty parameter $d$, whose tuning depends on the block production ratio that a designer wants to obtain.
Therefore we do not discuss the choice of $d$ but we refer to the BTC implementation \cite{BTCpow} or to more structured models such as personalized difficulty adjustments \cite{PersD}.
Our goal is to produce a prime number of the prescribed size and satisfying the following properties.

\footnotesize \textrm{EXCEPTIONALITY PROPERTIES}\\
\normalsize \fbox{\begin{minipage}{35.8em}
\begin{enumerate}
\item $p$ is not a Crandall prime \cite{Crandall}, i.e. not of the from $2^k - c$ for a relatively small and positive integer $c$.
\item $p$ is neither a Generalized Mersenne prime \cite{Solinas} nor a More Generalized Mersenne prime \cite{MoreMersenne}, i.e. it may not be written as $p(m)$ for some integer $m$ and polynomial $p$ with very small coefficients and number of monomials.
\item $p$ is not Montgomery-friendly \cite{Mont1, Mont2, Mont3}, i.e. it may not be obtained as $2^{\alpha}(2^{\beta}~-~\gamma)~-~1$ for small positive integers $\alpha, \beta, \gamma$.
\end{enumerate}
\end{minipage}}

\smallskip
Given the difficulty parameter $d$ and the hash of the previous header $h$, we propose the generation of such a prime number $p$ as follows.

\begin{lstlisting}
function p_Gen(d, h)
  repeat
    h = $\H$(h)
    p = NextPrime(h mod $2^{2d}$)
  until p satisfies exceptionality properties
  return p
\end{lstlisting}

\item Generating $E$

We aim at generating pseudorandom elliptic curves for which no efficient attacks are currently known, 
i.e. satisfying the following properties.

\footnotesize \textrm{SECURITY PROPERTIES}\\
\normalsize \fbox{\begin{minipage}{35.8em}
\begin{enumerate}
\item The number of points of $E$ is prime and different from $p$.
\item The \emph{embedding degree} $B$ is greater than 20, i.e. $|E| \nmid p^B-1$ for every $1 \leq B \leq 20$.
\item Let $D$ be the \emph{CM field discriminant}, defined as
\begin{equation*}
D = \begin{cases}
\Delta &\text{if } \Delta \equiv 1 \bmod 4,\\
4\Delta &\text{otherwise},
\end{cases} \hspace{0.8cm} \Delta = \text{SquareFreePart}(t^2-4p),
\end{equation*}
where $t$ is the trace of $E$. Then we require $D > 2^{40}$.
\end{enumerate}
\end{minipage}}

\smallskip
Let $h$ be the previous block header, we suggest to generate the curve as follows.

\vspace{0.5cm}
\begin{lstlisting}
function E_Gen(p, h)
  i = 0
  repeat
    i = i + 1
    $A_E$ = $\H$(h + i)
    $B_E$ =  $\H$($A_E$)
    E defined by $y^2 = x^3 + A_E x + B_E$ over $\mathbb{F}_p$
  until E is an EC satisfying security properties
  return E
\end{lstlisting}

\item Generating $P$

The base point we prescribe for an EB and its subsequent epoch is
\begin{equation*}
P = \verb|P_Gen|(\H(p\ ||\ A_E\ ||\ B_E), E).
\end{equation*}
\end{itemize}

The new epoch parameters are manufactured before the PoW production, which therefore depends on them.

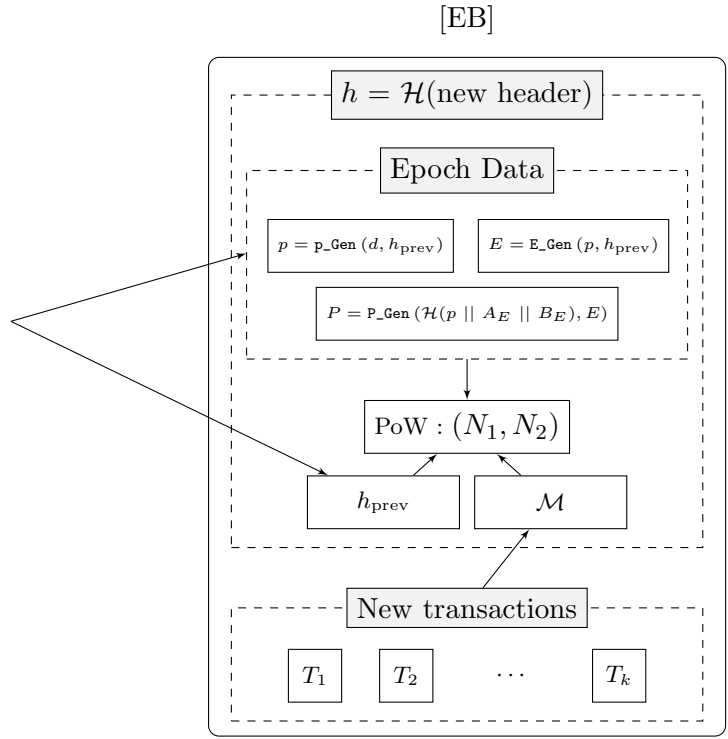
\begin{center}
\begin{tikzpicture}
\node (start) [rectangle, rounded corners, minimum width=6.8cm, minimum height=9cm, text centered, draw=black, fill=white!30] { };
\node [above of = start, node distance=+5.0cm] {[EB]};

\draw [dashed] (-2.9,3) rectangle ++(5.8,-2.5);
\node [draw = black, rectangle, above of = start, node distance=+2cm, fill=gray!10] (EpoData) at (0,1) {Epoch Data};

\node [draw = black, rectangle, minimum width=1.9cm, minimum height=0.7cm, above of = start, node distance=0.9cm] (Prev) at (-1.4,1.1) {\tiny $p = \verb|p_Gen|\,(d, h_{\text{prev}})$};
\node (pEP) [draw = black, rectangle, minimum width=1.9cm, minimum height=0.7cm, right of = Prev, node distance=2.8cm] {\tiny $E = \verb|E_Gen|\,(p, h_{\text{prev}})$};
\node (pEP) [draw = black, rectangle, minimum width=2cm, minimum height=0.7cm, below of = Prev, node distance=0.9cm, xshift = 1.4cm] {\tiny $P = \verb|P_Gen|\,(\H(p\ ||\ A_E\ ||\ B_E), E)$};

\draw [dashed] (-3.1,4.0) rectangle ++(6.2,-6.0);
\node (NewH) [draw = black, rectangle, above of = start, node distance=1.8cm, fill=gray!10] at (0,2.2) {$h$ = $\mathcal{H}$(new header)};

\node (PoW) [draw = black, rectangle, minimum width=2cm, minimum height=0.7cm, above of = start, node distance=1cm] at (0,-1.4) {\footnotesize PoW : \normalsize $(N_1,N_2)$};
\node (Prev) [draw = black, rectangle, minimum width=2cm, minimum height=0.7cm, above of = start, node distance=1cm] at (-1.1,-2.4) {\footnotesize $h_{\text{prev}}$};
\node (Merkle) [draw = black, rectangle, minimum width=2cm, minimum height=0.7cm, right of = Prev, node distance=2.2cm] {\footnotesize $\mathcal{M}$};

\draw [dashed] (-3.1,-2.8) rectangle ++(6.2,-1.5);
\node (NewT) [draw = black, rectangle, below of = start, node distance=1.2cm, fill=gray!10] at (0,-1.6) {New transactions};
\node (T1) [draw = black, rectangle, minimum width=0.7cm, minimum height=0.7cm, below of = (NewT), node distance=2.5cm] at (-2,-1.2) {\footnotesize $T_1$};
\node (T2) [draw = black, rectangle, minimum width=0.7cm, minimum height=0.7cm, below of = (NewT), node distance=2.5cm, node distance=2.5cm] at (-0.8,-1.2) {\footnotesize $T_2$};
\node (T3)[rectangle, minimum width=0.7cm, minimum height=0.7cm, below of = (NewT), node distance=2.5cm] at (0.6,-1.2) {\footnotesize $\cdots$};
\node [draw = black, rectangle, minimum width=0.7cm, minimum height=0.7cm, below of = (NewT), node distance=2.5cm] at (2,-1.2) {\footnotesize $T_k$};

\path [line] (0,0.5) -- node [midway,left] {} (PoW);
\path [line] (Merkle) -- node [midway,left] {} (PoW);
\path [line] (Prev) -- node [midway,left] {} (PoW);
\path [line] (NewT) -- node [midway,left] {} (Merkle);
\path [line] (-6,1) -- node [midway,above,xshift=-6.5mm] {} (Prev);
\path [line] (-6,1) -- node [midway,above,xshift=-6.5mm] {} (-2.9,1.9);
\end{tikzpicture}
\end{center} 

Despite the verification of SBs is extremely fast, EBs are slower to be checked since verifiers need to test that all the curve parameters involved have been properly constructed, running several types of mathematical algorithms such as primality testing, finite fields operations and points counting.

\section{Method discussion} \label{sec:Discussion}

Here we discuss motivation and advantages of the presented choices.

First, this PoW model involves many different mathematical algorithms of wide interest, for which this blockchain may represent a concrete research propellant. Furthermore, it might also provides a public collection of cryptographically secure elliptic curves of moderate size.

Apart from its scientific usefulness, it conveys many desirable security properties. The challenges involved do not rely on a given curve of questionable provenance but on the \emph{generic} difficulty of the ECDLP, which is much more fair to be trusted.
Thus, we find it aims at embracing the decentralization ideals that lead to cryptocurrencies creation: even the mathematical objects involved are publicly manufactured, no trust is required even in the authors or the proposing entities.

The existence of different types of block in blockchains has become common, as it is considered suitable for tackling the problem of scalability \cite{SurveySharding}.

As for blocks forgery, we point out that both SBs and EBs comprise a PoW which depends on the entire block, together with the previous one. This means that any counterfeit in any position of the chain results into an incorrect final block, which may be easily detected from the network.

Moreover, it is hard to conceive shortcuts for the PoW production: for a given difficulty parameter $d$ we expect a $d$-bits secutity of the \emph{general} ECDPL by using $p \approx 2^{2d}$, unless attacks outperforming Pollard's rho are discovered.
Moreover, common base field operations speed ups are avoided by making use of not-exceptional primes, ensuring a fair and general problem to be solved equally for every miner.
In fact, neither specific algorithms nor dedicated hardware may be used for solving such a general problem, of which easy cases are carefully avoided.
Also, the constructed curves fulfil the known security criteria \cite{SecCrit}:
\begin{itemize}
\item working over prime fields avoids Weil-descent attacks;
\item searching for curves of prime order prevents from Polig-Hellman attacks;
\item since $p \neq |E|$ the curves are not anomalous so Smart, Semaev, Satoh-Araki attacks do not apply;
\item the embedding degree we suggest is greater than 20 as required by SEC1 \cite{SEC1}, which prevents pairing attacks such as Menezes-Okamoto-Vanstone (based on Weil Pairing) and Frey-R\"uck (based on Tate-Lichtenbaum Pairing);
\item attacks to curves with low CM discriminant are prevented by requiring it higher than $2^{40}$, as for Brainpool Standard Curves \cite{Brainpool}.
\end{itemize}

\begin{thm}
Let us assume that the current epoch is endowed with the curve $E$ and its base-point $P$.
Let $\sigma$ be a deterministic digital signature algorithm and $\mathbb{M}$ be a proposing miner with fixed signing key $k$.
If $\mathbb{M}$ exhibits a valid block, then it has solved at least one generic instance of ECDLP on $E$.






\end{thm}

\begin{proof}

By definition of our PoW, the given block is valid if and only if it contains $(N_1,N_2)$ such that

\begin{equation*}
\begin{cases}
    Q_1 = \verb|P_Gen|(\H(\sigma_k( h_{\text{prev}}) ), E) = N_1 \cdot P,\\
    Q_2 = \verb|P_Gen|(\H(\mathcal{M}), E) = N_2 \cdot P.
\end{cases}
\end{equation*}

Since $E$ and $P$ are determined by the epoch and $h_{\text{prev}}$ is determined by the previous block, the proposing miner has no control on them.
Moreover, $\sigma$ is deterministic and $k$ is fixed, so $Q_1 = \verb|P_Gen|(\H(\sigma_k( h_{\text{prev}} )), E)$ cannot be influenced by the miner.
Therefore the miner must solve $Q_1 = N_1 \cdot P$.

\end{proof}

\begin{rmk}
Even if in the previous proof we do not consider the equation $Q_2 = N_2 \cdot P$, we believe that it adds extra security. Indeed, it is unlikely that a miner can avoid solving $Q_2 = N_2 \cdot P$ unless $\mathbb{M}$ has computed a multiple of $P$ with any $m$, $\bar Q=m P$,  and solved the hash preimage equation $\H(\mathcal{M})=x$, where $x$ is the x-coordinate of $\bar Q$.
\end{rmk}

\endproof

Besides security, the curves we propose are \emph{fully rigid} as defined in \cite{SecCrit}: their construction is entirely explained in terms of the previous block, which cannot be controlled by a malicious actor since there is no room for miner choices (such as nonces).
Even assuming that the transactions of the previous block might be chosen \emph{ad hoc}, an attacker who wants to impose a particular curve during the next epoch has to brute-force invert the hash $\H$ at the cost of one ECDLP solution for each attempt, until a desired hash digest is obtained, within the time needed for the entire network to solve a single ECDLP. We consider this scenario unachievable under realistic assumptions.

As regards the difference between EBs and SBs, we point out that the bulk of miner's work consists of the ECDLP solution: we expect good parameters to be generated in EBs in a time which is linear in the difficulty parameter \cite{ProbPrime} whereas the asymptotic difficulty of ECDLP solution is exponential in it.

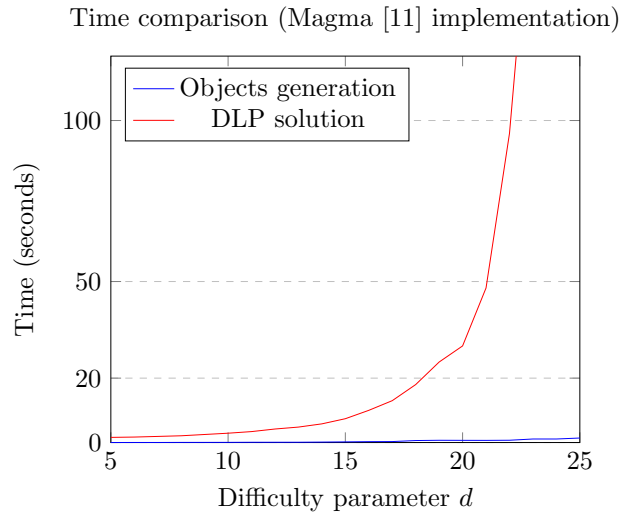
\begin{center}
\begin{tikzpicture}[scale=0.9]
\begin{axis}[
    title={Time comparison (Magma \cite{Magma} implementation)},
    xlabel={Difficulty parameter $d$},
    ylabel={Time (seconds)},
    xmin=5, xmax=25,
    ymin=0, ymax=120,
    xtick={5,10,15,20,25},
    ytick={0,20,50,100,200,500,700,1000},
    legend pos=north west,
    ymajorgrids=true,
    grid style=dashed,
]
 
\addplot[
    color=blue,
    ]
    coordinates {
    (5,0.007)(6,0.018)(7,0.018)(8,0.039)(9,0.042)(10,0.053)(11,0.082)(12,0.085)(13,0.092)(14,0.14)(15,0.20)(16,0.27)(17,0.30)(18,0.62)(19,0.70)(20,0.68)(21,0.67)(22,0.71)(23,1.1)(24,1.1)(25,1.4)
    };
    \addlegendentry{Objects generation}
 
 \addplot[
    color=red,
    ]
    coordinates {
    (5,1.6)(6,1.7)(7,1.9)(8,2.1)(9,2.5)(10,2.9)(11,3.4)(12,4.2)(13,4.8)(14,5.8)(15,7.4)(16,10)(17,13)(18,18)(19,25)(20,30)(21,48)(22,96)(23,180)(24,350)(25,820)
    };
    \addlegendentry{DLP solution}
    
\end{axis}
\end{tikzpicture}
\end{center}

Since the curves creation appears not to be computationally demanding when compared to the actual PoW, then lazy miners do not have any substantial advantage in skipping it.

\section{Conclusion} \label{sec:Conclusion}

We have proposed a new PoW-based blockchain model based on \emph{general} ECDLP, highlighting the desirable properties that such a scheme provides in terms of scientific relevance, security and pure decentralization ideals.

The past proposals \cite{DLOG, CryptoTool} have the high merit of introducing ECDLP as a problem whose solution provides consensus, but we felt compelled to remove the suspiscious choice of the curve serving as a common battlefield for miners.

It may be interesting to produce an actual implementation of the proposed scheme, obtaining practical time measurments and efficiency considerations.
A subsequent engaging project might address the resistance of such a protocol to the known attacks under real-world assumptions, comparing the obtained results with outcomes of existing cryptocurrencies.
Further studies may also be carried on other types of curve models, such as Edwards or Montgomery curves. Even though this is likely to improve the overall performance of this scheme, it should be observed that it contrasts with our declared intention of making use of \emph{general} objects.

Finally, different types of PoW might be conceived in a similar fashion, possibly employing problems which are thought to resist even to quantum attacks.

\section*{Aknowledgments}
The results presented here have been carried on within the EU-ESF activities, call ”PON Ricerca e Innovazione 2014-2020”, project “Distributed Ledgers for Secure Open Communities”. We thank the Quadrans Foundation for its support.

\end{document}